\documentstyle[preprint,aps,epsfig]{revtex}
\begin{document}

\preprint{\vbox{\baselineskip16pt
\hbox{SNUTP 99-030}
}}
\title{$\Lambda_b\to\Sigma_c^{(*)}\pi l{\bar\nu}$ in the heavy quark, chiral,
       and large $N_c$ limits}

\author{
  Jong-Phil Lee$^a$\footnote{Email address:~jplee@phya.snu.ac.kr},
  Chun Liu$^{b,c}$\footnote{Email address:~liuc@itp.ac.cn}, and
  H.~S. Song$^a$\footnote{Email address:~hssong@physs.snu.ac.kr}}
\address{
  $^a$Department of Physics and Center for Theoretical Physics,\\
  Seoul National University, Seoul 151-742, Korea}
\address{
  $^b$Institute of Theoretical Physics, Chinese Academy of Sciences,\\ 
  P.O. Box 2735, Beijing 100080, China\\
  $^c$Korea Institute for Advanced Study,\\
  207-43 Cheongryangri-dong, Dongdaemun-gu, Seoul 130-012, Korea}
\maketitle

\begin{abstract}

The decay of $\Lambda_b\to\Sigma_c^{(*)}\pi l{\bar\nu}$ processes are 
calculated in the framework of heavy baryon chiral perturbation theory with 
help of the $1/N_c$ expansion.  The decay rate distributions over 
$v\cdot v^{\prime}$ where $v$ and $v^{\prime}$ are the 4-vector velocities of 
$\Lambda_b$ and $\Sigma_c^{(*)}$ respectively, and over the pion and lepton 
energies are given.  The decay branching ratios amount to about 
$1.21(0.971)\%$ for $\Lambda_b\to\Sigma_c\pi l{\bar\nu}$ and $0.798(0.640)\%$ 
for $\Lambda_b\to\Sigma_c^*\pi l{\bar\nu}$ by using the baryonic Isgur-Wise 
function predicted by QCD sum rule (large $N_c$ QCD).

\end{abstract}


\section{Introduction}

Heavy baryons provide a testing ground to the standard model (SM).
Experiments are accumulating more and more data on heavy baryons \cite{Caso}.
For the theoretical calculation, the task is to apply model-independent 
methods of the nonperturbative QCD.
For the heavy hadrons containing a single heavy quark, the heavy quark
effective theory (HQET) \cite{review} can be applied successfully.
After combining the chiral symmetry, the low energy effective theory of HQET,
the heavy hadron chiral Lagrangian (HHCL) \cite{Wise} extends 
the application area further.
Recently, the $1/N_c$ expansion approach \cite{Dashen} incorporates in more
understanding of QCD to the baryon sector.
\par
The heavy baryon decays $\Lambda_b\to\Sigma_c^{(*)}\pi l{\bar\nu}$
\cite{Cho}
are interesting from the point of view of the above mentioned approaches.
In the heavy quark limit, the most interesting weak transitions
$\Lambda_b\to\Lambda_c$ and $\Sigma_b^{(*)}\to\Sigma_c^{(*)}$ are
simplified greatly.
The numbers of the independent form factors are reduced to one and two,
respectively in these two transitions.
The decays $\Lambda_b\to\Sigma_c^{(*)}l{\bar\nu}$ \cite{add} however, are 
predicted to be both $1/m_Q$ and $1/N_c$ suppressed.
They provide a test to the HQET.
The decays $\Lambda_b\to\Sigma_c^{(*)}\pi l{\bar\nu}$, on the other hand,
are neither $1/m_Q$ nor $1/N_c$ suppressed.
They may be the dominant modes of the decays $\Lambda_b\to\Sigma_c^{(*)} X$,
although they are phase space suppressed.
We will calculate them in the heavy quark limit.
The calculation involves in the HHCL by assuming the pion being soft.
It also needs the three form factors of 
$\Lambda_b\to\Lambda_c$ and $\Sigma_b^{(*)}\to\Sigma_c^{(*)}$.
In the large $N_c$ limit, there is a light quark spin-flavor symmetry 
in the baryon sector.
This leads to the fact that only one form factor is independent 
among the three \cite{Chow}.
It is argued that this result is true until to the order of
$1/N_c^2$ \cite{Liu}.
\par

This paper is organized as follows.
In Sec. II, after a brief review of the HHCL, the invariant matrix 
elements of the decays are calculated in this framework with the help of  
the $1/N_c$ expansion.  Sec. III describes necessary kinematics for the phase 
space integration and presents the numerical results.  The discussion and 
summary are made in Sec. IV.

\section{Dynamics}

The dynamics of the decays $\Lambda_b\to\Sigma_c^{(*)}\pi l{\bar\nu}$ is 
shown in Fig. 1, which includes both the strong and the weak interactions.  The 
strong interaction is described by the HHCL in the case of the pion being soft.  
The chiral perturbation theory is the low energy effective theory of QCD for 
the light pseudoscalar mesons.  These mesons are regarded as the Goldstone 
bosons due to the spontaneous chiral symmetry breaking 
${\rm SU(3)}_L\times{\rm SU(3)}_R\to {\rm SU(3)}_V$ in QCD.
The eight Goldstone boson fields, namely the pion, the 
kaon and the $\eta$-meson fields, can be represented in the nonlinear 
exponential form \cite{georgi2}
\begin{equation}
\Sigma=\exp(2i\tilde{\pi}/f)~,
\end{equation}
where
\begin{equation}
\tilde{\pi}=\frac{1}{\sqrt{2}}\left(
 \begin{array}{ccc}
 \sqrt{\frac{1}{2}}\pi^0+\sqrt{\frac{1}{6}}\eta & \pi^+ & K^+ \\
 \pi^- & -\sqrt{\frac{1}{2}}\pi^0+\sqrt{\frac{1}{6}}\eta & K^0 \\
 K^- & {\bar K}^0 & -\sqrt{\frac{2}{3}}\eta 
 \end{array}\right)~,
\end{equation}
and $f$ is the pion decay constant, $f= 93 ~{\rm MeV}$.
It is convenient to introduce 
\begin{equation}
\xi\equiv\sqrt{\Sigma}=\exp(i\tilde{\pi}/f)~.
\end{equation}
$\Sigma$ and $\xi$ transform under the chiral symmetry as
\begin{equation}
 \Sigma\rightarrow L\Sigma R^\dagger~,~~~
 \xi\rightarrow L\xi U^\dagger=U\xi R^\dagger~,
\label{transform}
\end{equation}
where $L(R)\in{\rm SU(3)}_{L(R)}$ and the special unitary matrix $U$ is a
complicated nonlinear function of $L$, $R$, and $\tilde{\pi}$.
To the leading order of the derivative expansion, the effective Lagrangian 
for these Goldstone bosons is 
\begin{equation}
{\cal L}_0 = \frac{1}{4}f^2{\rm Tr}(\partial^\mu\Sigma^\dagger
             \partial_\mu\Sigma)~.
\label{lagrangian}
\end{equation}

Heavy baryons can be included in the chiral Lagrangian formalism.  
The ground state heavy baryons can be classified into two categories.
One is the ${\rm SU(3)}_V$ sextet field $S^{ij}_{\mu}$ which describes the 
heavy baryons whose light quarks have one unit of angular momentum.  The 
resulting total angular momentum of these baryons is either 1/2 or 3/2.  
$S_{\mu}^{ij}$ is written as \cite{georgi}
\begin{equation}
S^{ij}_\mu(v)=
  \sqrt{\frac{1}{3}}(\gamma_\mu+v_\mu)\gamma_5
  \frac{1+v\hspace{-2mm}/}{2}B^{ij}+\frac{1+v\hspace{-2mm}/}{2}B^{*ij}_\mu~,
\end{equation}
where $v$ is the 4-velocity of the heavy baryon.  $B^{ij}$ is the Dirac field 
and $B^{*ij}_\mu$ is its spin-3/2 counterpart.
$B^{ij}$ represents the following $3\times 3$ matrix 
\begin{equation}
B_{ij}=\left(\begin{array}{ccc}
           \Sigma^{+1}_Q & \frac{1}{\sqrt{2}}\Sigma^0_Q 
	        & \frac{1}{\sqrt{2}}\Xi^{\prime+\frac{1}{2}}_Q \\
           \frac{1}{\sqrt{2}}\Sigma^0_Q & \Sigma^{-1}_Q 
	        & \frac{1}{\sqrt{2}}\Xi^{\prime-\frac{1}{2}}_Q \\
           \frac{1}{\sqrt{2}}\Xi^{\prime+\frac{1}{2}}_Q 
	        & \frac{1}{\sqrt{2}}\Xi^{\prime-\frac{1}{2}}_Q
	        & \Omega_Q 
	   \end{array}\right)~,
\end{equation}
and $B^{*ij}_\mu$ has a similar form.  The superscripts denote the charge and 
the isospin.  Under the chiral symmetry, $S^{ij}_\mu$ transforms as
\begin{equation}
S^{ij}_\mu\rightarrow
  U^i_{i^\prime}S^{i^\prime j^\prime}_\mu U^j_{j^\prime}~.
\end{equation}
Under the heavy quark spin ${\rm SU(2)}_v$ symmetry,
\begin{equation}
 S_\mu\rightarrow\exp^{{\rm i}{\bf\epsilon}\cdot{\bf S}_v}S_\mu~,
\label{SU(2)}
\end{equation}
where ${\bf S}_v$ is the ${\rm SU(2)}$ generator and ${\bf\epsilon}$ is an
infinitesimal parameter.  The other component in the classification 
is the ${\rm SU(3)}_V$ anti-triplet field $T_i$ corresponding to the baryons 
whose light quarks have zero angular momentum.  The resulting total angular 
momentum of the baryons is 1/2.  It can be expressed as
\begin{equation}
T_i(v)=\frac{1+v\hspace{-2mm}/}{2}B_i~,
\end{equation}
where the Dirac field $B_i$ stands for 
\begin{equation}
B_1=\Xi^{-\frac{1}{2}}_Q~,~~~B_2=-\Xi^{+\frac{1}{2}}_Q~,~~~B_3=\Lambda_Q~.
\end{equation}
Under the chiral symmetry, it transforms as
\begin{equation}
T_i\rightarrow T_j (U^\dagger)^j_i~.
\end{equation}
Under the ${\rm SU(2)}_v$, 
it transforms just like (\ref{SU(2)}).\par

To the leading order of the derivative expansion, the effective Lagrangian of 
the strong interaction of $S^{ij}_\mu$ and $T_i$ with the Goldstone bosons, 
which is invariant under the chiral symmetry and the heavy 
quark symmetry is \cite{Cho}
\begin{eqnarray}
{\cal L}_{int}&=&\sum_{\rm heavy~quarks}\big\{
   -{\rm i}{\bar S}^\mu_{ij}v\cdot{\cal D}S^{ij}_\mu
   +{\rm i}{\bar T}^i v\cdot{\cal D}T_i
   +{\rm i}g_2\epsilon_{\mu\nu\sigma\lambda}
       {\bar S}^\mu_{ik}v^\nu(A^\sigma)^i_j(S^\lambda)^{jk}\nonumber\\
  & & +g_3[\epsilon_{ijk}{\bar T}^i(A^\mu)^j_lS^{kl}_\mu
       +\epsilon^{ijk}{\bar S}^\mu_{kl}(A_\mu)^l_j T_i]
   +\Delta \bar{S}^{\mu}_{ij} S^{ij}_{\mu}
  \big\}~,
\label{int}
\end{eqnarray}
where $g_i$'s ($i=1,2,3$) denote the coupling constants and the covariant 
derivatives are
\begin{eqnarray}
{\cal D}^\mu S^{ij}_\nu &=& 
   \partial^\mu S^{ij}_\nu+(V^\mu)^i_k S^{kj}_\nu+(V^\mu)^j_k S^{ik}_\nu~,
   \nonumber\\
{\cal D}^\mu T_i &=& \partial^\mu T_i - T_j(V^\mu)^j_i~,
\end{eqnarray}
and
\begin{eqnarray}
{\bf V}^\mu &=& 
  \frac{1}{2}(\xi^\dagger\partial^\mu\xi+\xi\partial^\mu\xi^\dagger)~,
  \nonumber\\
{\bf A}^\mu &=&
  \frac{1}{2}{\rm i}(\xi^\dagger\partial^\mu\xi-\xi\partial^\mu\xi^\dagger)~.
\end{eqnarray}
The mass difference between the sextet and the anti-triplet baryons is
\begin{equation}
\Delta=m_S-m_T~.
\end{equation}
We have neglected the mass differences between the $\Sigma_Q$ and $\Sigma_Q^*$
which is $1/m_Q$ suppressed.
\par

The weak interaction of $b\to cl\nu$ is described by
\begin{equation}
{\cal L}_{\rm weak}=\frac{4G_F}{\sqrt{2}}V_{cb}
    \sum_l({\bar l}\gamma^\mu\frac{1-\gamma_5}{2}\nu_l)
    ({\bar c}\gamma_\mu\frac{1-\gamma_5}{2}b)~,
\end{equation}
and the hadronic current in the heavy quark limit can be re-expressed by the 
matter fields as 
\begin{eqnarray}
{\bar c}\gamma_\mu\frac{1-\gamma_5}{2}b &\to& C_{cb}\big\{
  [-g_{\alpha\beta}\eta_1(v\cdot v^\prime)
     +v_\alpha v^\prime_\beta\eta_2(v\cdot v^\prime)]
  {\bar S}^\alpha_c(v^\prime)\gamma_\mu\frac{1-\gamma_5}{2} S^\beta_b(v)
  \nonumber\\
  & &+\eta(v\cdot v^\prime){\bar T}_c(v^\prime)\gamma_\mu
     \frac{1-\gamma_5}{2}T_b(v)\big\}~,
\end{eqnarray}
where $\eta$, $\eta_1$, and $\eta_2$ are the Isgur-Wise functions.  And 
with the definition $\omega=v\cdot v'$, 
the perturbative QCD correction factor $C_{cb}$ is 
\begin{eqnarray}
C_{cb}(v\cdot v^\prime)&=&\Bigg[
  \frac{\alpha_s(m_b)}{\alpha_s(m_c)}\Bigg]^{-\frac{6}{25}}
  \Bigg[\frac{\alpha_s(m_c)}{\alpha_s(\mu)}\Bigg]^{a_L(v\cdot v^\prime)}~,
  \nonumber\\
a_L(w)&=&\frac{8}{27}\Big[\frac{w}{\sqrt{w^2-1}}
  \ln\big(w+\sqrt{w^2-1}\Big)-1\big]~.
\end{eqnarray}
In the large $N_c$ limit, the Isgur-Wise functions have the following 
relations \cite{Chow,Liu}
\begin{equation}
\eta(w)=\eta_1(w),~~~\eta_2(w)=\frac{\eta(w)}{w+1}~.
\end{equation}
\par

In calculating Fig. 1, the spinor sum relations for the
Dirac and Rarita-Schwinger types are used,
\begin{eqnarray}
\sum_s u(v,s){\bar u}(v,s) &=& \frac{1+v\hspace{-2mm}/}{2}~,\nonumber\\
\sum_s {\cal U}^\mu(v,s){\bar {\cal U}}^\nu(v,s) &=&
   [-g^{\mu\nu}+v^\mu v^\nu
    +\frac{1}{3}(\gamma^\mu+v^\mu)(\gamma^\nu-v^\nu)]\cdot
    \frac{1+v\hspace{-2mm}/}{2}~,
\end{eqnarray}
where $u(v,s)$ is the Dirac spinor and ${\cal U}(v,s)$ is the Rarita-Schwinger 
spinor.
We obtain the squared invariant matrix elements as
\begin{equation}
\frac{1}{2}\sum_{\rm spin}|{\cal M}(\Lambda_b(v)\to\Sigma_c(v')\pi l\nu)|^2
=\frac{16G_F^2}{3f^2}g_3^2|V_{cb}|^2C_{cb}^2\eta^2
\frac{p_l\cdot v^\prime~ p_\nu\cdot v ~A}
 {(p_\pi\cdot v+\Delta)^2(p_\pi\cdot v^\prime+\Delta)^2}~,
 \label{ampa}
\end{equation}
where
\begin{eqnarray}
A&=&2(p_\pi\cdot v)^3 p_\pi\cdot v^\prime
       +2(p_\pi\cdot v)^3 \Delta
      +4(p_\pi\cdot v)^2 (p_\pi\cdot v^\prime)^2
     +6(p_\pi\cdot v)^2 p_\pi\cdot v^\prime\Delta\nonumber\\
      &&+(p_\pi\cdot v)^2\Delta^2w
      +3(p_\pi\cdot v)^2\Delta^2
      -(p_\pi\cdot v)^2m_\pi^2w
      -(p_\pi\cdot v)^2m_\pi^2\nonumber\\
      &&+2p_\pi\cdot v (p_\pi\cdot v^\prime)^3
      +6p_\pi\cdot v (p_\pi\cdot v^\prime)^2\Delta
      -2p_\pi\cdot v~ p_\pi\cdot v^\prime\Delta^2w
      +2p_\pi\cdot v~ p_\pi\cdot v^\prime\Delta^2\nonumber\\
      &&-2p_\pi\cdot ~v p_\pi\cdot v^\prime m_\pi^2w
      -2p_\pi\cdot v ~p_\pi\cdot v^\prime m_\pi^2
     -4p_\pi\cdot v\Delta m_\pi^2 w
      -4p_\pi\cdot v\Delta m_\pi^2\nonumber\\
     &&+2(p_\pi\cdot v^\prime)^3\Delta
     +(p_\pi\cdot v^\prime)^2\Delta^2w
     +3(p_\pi\cdot v^\prime)^2\Delta^2
     -(p_\pi\cdot v^\prime)^2 m_\pi^2 w\nonumber\\
     &&-(p_\pi\cdot v^\prime)^2 m_\pi^2
     -4p_\pi\cdot v^\prime\Delta m_\pi^2 w
     -4p_\pi\cdot v^\prime\Delta m_\pi^2
      -4\Delta^2 m_\pi^2 w
     -4\Delta^2 m_\pi^2)~, \nonumber\\
 \label{amp1}
\end{eqnarray}
and
\begin{equation}
\frac{1}{2}\sum_{\rm spin}|{\cal M}(\Lambda_b(v)\to\Sigma_c^*(v')\pi l\nu)|^2
\nonumber\\[3mm]
=\frac{16G_F^2}{3f^2}g_3^2|V_{cb}|^2C_{cb}^2\eta^2
\frac{B}{(p_\pi\cdot v+\Delta)^2(p_\pi\cdot v^\prime+\Delta)^2}~,
 \label{ampb}
\end{equation}
where
\begin{eqnarray}
B&=&
-p_l\cdot p_\nu~(p_\pi\cdot v)^2~(p_\pi\cdot v^\prime)^2w
-p_l\cdot p_\nu~(p_\pi\cdot v)^2~(p_\pi\cdot v^\prime)^2
+p_l\cdot p_\nu~(p_\pi\cdot v)^2~m_\pi^2w\nonumber\\
&&+p_l\cdot p_\nu~(p_\pi\cdot v)^2~m_\pi^2 
-2p_l\cdot p_\nu~p_\pi\cdot v~(p_\pi\cdot v^\prime)^2~\Delta w
-2p_l\cdot p_\nu~p_\pi\cdot v~(p_\pi\cdot v^\prime)^2~\Delta\nonumber\\
&&+2p_l\cdot p_\nu~p_\pi\cdot v~\Delta m_\pi^2w
+2p_l\cdot p_\nu~p_\pi\cdot v~\Delta m_\pi^2
-p_l\cdot p_\nu~(p_\pi\cdot v^\prime)^2\Delta^2w\nonumber\\
&&-p_l\cdot p_\nu~(p_\pi\cdot v^\prime)^2\Delta^2
+p_l\cdot p_\nu~\Delta^2m_\pi^2w
+p_l\cdot p_\nu~\Delta^2m_\pi^2\nonumber\\
&&+4p_l\cdot v^\prime~p_\nu\cdot v~(p_\pi\cdot v)^3~p_\pi\cdot v^\prime
+4p_l\cdot v^\prime~p_\nu\cdot v~(p_\pi\cdot v)^3~\Delta\nonumber\\
&&-4p_l\cdot v^\prime~p_\nu\cdot v~(p_\pi\cdot v)^2~(p_\pi\cdot v^\prime)^2w
+4p_l\cdot v^\prime~p_\nu\cdot v~(p_\pi\cdot v)^2~(p_\pi\cdot v^\prime)^2\nonumber\\
&&+12p_l\cdot v^\prime~p_\nu\cdot v~(p_\pi\cdot v)^2~p_\pi\cdot v^\prime\Delta
+2p_l\cdot v^\prime~p_\nu\cdot v~(p_\pi\cdot v)^2~\Delta^2w\nonumber\\
&&+6p_l\cdot v^\prime~p_\nu\cdot v~(p_\pi\cdot v)^2~\Delta^2
+2p_l\cdot v^\prime~p_\nu\cdot v~(p_\pi\cdot v)^2~m_\pi^2w\nonumber\\
&&+2p_l\cdot v^\prime~p_\nu\cdot v~(p_\pi\cdot v)^2~m_\pi^2
+4p_l\cdot v^\prime~p_\nu\cdot v~p_\pi\cdot v~(p_\pi\cdot v^\prime)^3\nonumber\\
&&-8p_l\cdot v^\prime~p_\nu\cdot v~p_\pi\cdot v~(p_\pi\cdot v^\prime)^2~\Delta w
+4p_l\cdot v^\prime~p_\nu\cdot v~p_\pi\cdot v~(p_\pi\cdot v^\prime)^2~\Delta\nonumber\\
&&-4p_l\cdot v^\prime~p_\nu\cdot v~p_\pi\cdot v~p_\pi\cdot v^\prime~\Delta^2w
+4p_l\cdot v^\prime~p_\nu\cdot v~p_\pi\cdot v~p_\pi\cdot v^\prime~\Delta^2\nonumber\\
&&-4p_l\cdot v^\prime~p_\nu\cdot v~p_\pi\cdot v~p_\pi\cdot v^\prime~m_\pi^2w
-4p_l\cdot v^\prime~p_\nu\cdot v~p_\pi\cdot v~p_\pi\cdot v^\prime~m_\pi^2\nonumber\\
&&+4p_l\cdot v^\prime~p_\nu\cdot v~(p_\pi\cdot v^\prime)^3\Delta
-2p_l\cdot v^\prime~p_\nu\cdot v~(p_\pi\cdot v^\prime)^2\Delta^2w\nonumber\\
&&+2p_l\cdot v^\prime~p_\nu\cdot v~(p_\pi\cdot v^\prime)^2\Delta^2
-2p_l\cdot v^\prime~p_\nu\cdot v~(p_\pi\cdot v^\prime)^2m_\pi^2w
-2p_l\cdot v^\prime~p_\nu\cdot v~(p_\pi\cdot v^\prime)^2m_\pi^2\nonumber\\
&&-8p_l\cdot v^\prime~p_\nu\cdot v~p_\pi\cdot v^\prime~\Delta~m_\pi^2w
-8p_l\cdot v^\prime~p_\nu\cdot v~p_\pi\cdot v^\prime~\Delta~m_\pi^2
-4p_l\cdot v^\prime~p_\nu\cdot v~\Delta^2~m_\pi^2w\nonumber\\
&&-4p_l\cdot v^\prime~p_\nu\cdot v~\Delta^2~m_\pi^2~.
 \label{amp2}
\end{eqnarray}
In Eqs. (\ref{ampa}$-$\ref{amp2}), $p_\pi$, $p_l$ and $p_\nu$ are the momentums of
the pion, lepton and neutrino, respectively.  
\section{Kinematics and Results}

The kinematics of the four-body decay has been described in Ref. \cite{Cheng}.
It is shown in Fig. 2.  Under the definition of 
\begin{equation}
s_H\equiv (p_{\Sigma_c^{(*)}}+p_\pi)^2~,~~~s_L\equiv (p_l+p_\nu)^2~,
\end{equation}
the total decay rate is given by 
\begin{eqnarray}
d\Gamma &=&
 \frac{1}{2m_{\Lambda_b}}\frac{d^3p_{\Sigma_c^{(*)}}}{(2\pi)^3 2E_{\Sigma_c^{(*)}}}
 \frac{d^3p_l}{(2\pi)^3 2E_l}
 \frac{d^3p_\nu}{(2\pi)^3 2E_\nu}
 \frac{d^3p_\pi}{(2\pi)^3 2E_\pi}\nonumber\\
&& \times(2\pi)^4\delta^4(p_{\Lambda_b}-p_{\Sigma_c^{(*)}}-p_l-p_\nu-p_\pi)
 \Bigg(\frac{1}{2}\sum_{\rm spin}|{\cal M}|^2\Bigg)\nonumber\\
&=& \frac{1}{4(4\pi)^6 m_{\Lambda_b}^3}
    X\beta ds_H ds_L d\cos\theta d\cos\theta_l d\phi
 ~\Bigg(\frac{1}{2}\sum_{\rm spin}|{\cal M}|^2\Bigg)~,
\end{eqnarray}
where
\begin{eqnarray}
X&=&\frac{1}{2}\big(m_{\Lambda_b}^4+s_H^2+s_L^2-2m_{\Lambda_b}^2s_H-2s_Hs_L
    -2m_{\Lambda_b}^2s_L
   \big)^{\frac{1}{2}}
  = m_{\Lambda_b}\sqrt{(E_{\Sigma_c^{(*)}}+E_\pi)^2-s_H}~, \nonumber\\
\beta&=&\frac{1}{s_H}\big[(s_H+m_\pi^2-m_{\Sigma_c^{(*)}}^2)^2
       -4m_\pi^2s_H\big]^{\frac{1}{2}}~,
\end{eqnarray}
and $E_{\Sigma_c^{(*)}}$, $E_\pi$ are measured in $\Lambda_b$ rest frame.

To get the decay rate distributions, the following values are used,
\begin{eqnarray}
m_{\Lambda_b} &=& 5.64~{\rm GeV},~~~m_{\Sigma_c^{(*)}}=2.453~{\rm GeV},~~~
f=0.093~{\rm GeV},\nonumber\\
V_{cb} &=& 0.041,~~~\Lambda_{\rm QCD}=0.2~{\rm GeV},~~~\Delta=0.2~{\rm GeV}~.
\end{eqnarray}
The value of $g_3$ is $0.99$ \cite{cy}.  
We adopt two forms for $\eta(w)$.
One is a linear fitting of the result from QCD sum rule $\eta(w)=1-0.55(w-1)$
\cite{dai}; and
the other is an exponential function from large $N_c$ QCD 
$\eta(w)=0.99\exp[-1.3(w-1)]$ \cite{jenkins2}.\par

The numerical results of $\Lambda_b\to\Sigma_c \pi l{\bar\nu}$ and
$\Lambda_b\to\Sigma_c^* \pi l{\bar\nu}$ are given in Figs. 3 and 4, 
respectively.  The decay rate distributions on $\omega=v\cdot v'$, pion 
energy and lepton energy are shown.  The results of the two kinds of 
Isgur-Wise function are compared.

The total decay rates are obtained as 
\begin{eqnarray}
\Gamma(\Lambda_b\to\Sigma_c\pi l{\bar\nu})&=&
  6.45\times 10^{-15}~ (5.20\times 10^{-15})~{\rm GeV},\nonumber\\
\Gamma(\Lambda_b\to\Sigma_c^*\pi l{\bar\nu})&=&
  4.27\times 10^{-15}~ (3.43\times 10^{-15})~{\rm GeV},
\end{eqnarray}
and hence the branching ratios are
\begin{eqnarray}
{\cal B}(\Lambda_b\to\Sigma_c\pi l{\bar\nu})&=&1.21~(0.971)~\%~,\nonumber\\
{\cal B}(\Lambda_b\to\Sigma_c^*\pi l{\bar\nu})&=&0.798~(0.640)~\%~,
\end{eqnarray}
for the linear (exponential) $\eta(w)$.
Compared with the result of $\Lambda_b\to\Lambda_c l\nu$ \cite{jplee},
$\Gamma(\Lambda_b\to\Sigma_c\pi l \nu)/\Gamma(\Lambda_b\to\Lambda_c l\nu)\simeq 
10.5~(11.5)~\%$,
and 
$\Gamma(\Lambda_b\to\Sigma_c^*\pi l\nu)/\Gamma(\Lambda_b\to\Lambda_c l\nu)\simeq 
6.94.~(7.59)~\%$.
They are less dependent on the Isgur-Wise function because of the partial
cancellation of it in the ratios.

\section{Summary and Discussion}

We have calculated the decays $\Lambda_b\to\Sigma_c^{(*)}\pi l{\bar\nu}$ in the
heavy quark, chiral and large $N_c$ limits.  The differential decay spectra are 
presented.  The branching ratios are obtained as  
${\cal B}(\Lambda_b\to\Sigma_c\pi l{\bar\nu})=$ 1.21~(0.971)~\%, and 
${\cal B}(\Lambda_b\to\Sigma_c^*\pi l{\bar\nu})=$ 
0.798~(0.640)~\% with $\eta(w)$ from QCD sum rule (large $N_c$ QCD).

Let us discuss the accuracy of these approximations.  
First, consider the finite heavy quark mass correction.  We have taken 
$m_Q\to\infty$.  Generally the size of the $1/m_Q$ correction can be around 
$\Lambda_{\rm QCD}/m_c\simeq 20\%$.  The detailed consideration indicates that 
the correction maybe smaller.  In the 
calculation, there are two places where the heavy quark limit is applied.  One is 
in the weak transition where the quark current was replaced by the baryonic matrix
elements.  For the $\Lambda_b\to\Lambda_c$ transition, the $1/m_Q$ correction was 
calculated to be $11\%$ \cite{jplee}.  However, there is no estimation for the
$1/m_Q$ correction to the $\Sigma_b^{(*)}\to\Sigma_c^{(*)}$ transition.  It is 
experiential to guess that the correction in this case is also at the $10\%$ 
level.  The other place where the heavy quark limit was used is in the description 
of the strong interaction, namely in the HHCL.  The difference between the coupling 
$g_3$ for the bottom quark and that for the charm quark, and the difference between 
the masses of $\Sigma_Q$ and $\Sigma_Q^*$, have been neglected.  Theoretically these 
differences are both $1/m_Q$ and $1/N_c^2$ suppressed \cite{Dashen,jenkins}.  Note 
that practically $m_{\Sigma_c^*}-m_{\Sigma_c}\simeq 60$ MeV \cite{Caso} which is 
smaller than the pion mass.  Hence the correction in the second place is very 
small $\sim\frac{\Lambda_{\rm QCD}}{N_c^2m_c}\approx 2\%$.  From the 
above discussion, we expect that the $1/m_Q$ correction to our results is less 
than $20\%$, and is probably around $10\%$.  

Second, the validity of the soft pion limit should be addressed.  
The calculation is valid in the 
phase space region where the pion energy is smaller than the chiral symmetry 
breaking scale which is argued to be at $4\pi f\simeq1.2$ GeV 
\cite{georgi2}.  Therefore for a large part of phase space, the calculation 
is expected to be reliable.  However, the subleading terms in the derivative 
expansion in the effective Lagrangian have been omitted.  And because of the 
unknown coefficients of the subleading terms, it is still not possible to include 
them in the calculation.  The correction due to these terms is at the order of 
$v\cdot p_\pi/4\pi f$ or $v'\cdot p_\pi/4\pi f$ which is less than $30\%$ if the 
pion is really soft, say $E_\pi\leq 0.5$ GeV.  For the total decay rate, the 
results we have obtained maybe subject to larger correction.  

Finally, the $1/N_c$ expansion used in the calculation needs to be explicated.
The calculation needs the information of the three Isgur-Wise functions of the 
transitions $\Lambda_b\to\Lambda_c$ and $\Sigma_b^{(*)}\to\Sigma_c^{(*)}$.  In 
the large $N_c$ limit, the three form factors are related to each other 
\cite{Chow}.  These relations are argued to be valid to the order of $1/N_c^2$ 
\cite{Liu} due to 
the light quark spin-flavor symmetry in the baryon sector in the large $N_c$ 
limit.  The argument goes like that as in Ref. \cite{Dashen}.  It is the relation 
among the form factors, not the knowledge about the form factor itself, that we 
are considering.  Therefore the application of these relations causes  
$1/N_c^2\simeq 10\%$ uncertainty.  

Up to the above approximations, the calculation is in principle 
model-independent.  To get the final numerical results, the model result for 
the $\Lambda_b\to\Lambda_c$ Isgur-Wise function has to be used.  This step 
brings one more source of uncertainty.  According to the model results we 
adopted, the uncertainty is at $20\%$ level.

The decays $\Lambda_b\to\Sigma_c^{(*)}\pi l{\bar\nu}$ are indeed experimentally 
interesting according to our results.   The estimated branching ratios of the 
decays are not small.  The experiment has been studying the channel 
$\Lambda_b\to\Lambda_c l{\bar\nu}$ \cite{exp2}.  A lot of data on $\Lambda_b$ can
be expected to emerge in the near future from CDF, D0, HERA-B, and LHC-B and so on.  
With high statistics of the data, the decay channels 
$\Lambda_b\to\Sigma_c^{(*)}\pi l{\bar\nu}$ can be discovered.  

In the near future, more reliable theoretical results for the baryonic Isgur-Wise 
functions can be obtained, even for that appear in the subleading order of the 
heavy quark expansion, from lattice HQET or the other model-independent methods.  
That will give more accurate results of the decay distributions and decay rates.  
The uncertainties due to the heavy quark limit, large $N_c$ limit and the 
model-dependence can be reduced to $10\%$ or smaller.  In this case, the precise 
experimental measurement for $\Lambda_b\to\Sigma_c^{(*)}\pi l{\bar\nu}$ will give 
more information about the validity range of the HHCL.

\begin{center}
{\large\bf Acknowledgments}\\[10mm]
\end{center}\par

We would like to thank Hai-Yang Cheng for helpful discussions.
The work was supported in part by KOSEF through the SRC program of SNU CTP.
JPL and HSS also would like to acknowledge the support by the Korean Research 
Foundation through the 97-Sughak program and 1998-015-D00054.


\newpage
\begin{center}{\large\bf FIGURE CAPTIONS}\end{center}

\noindent
Fig.~1
Tree level diagrams for $\Lambda_b\to\Sigma_c^{(*)}\pi l\nu$.
Solid square(circle) indicates the weak(strong) interaction vertex.
\\
\vskip .3cm

\noindent
Fig.~2
The kinematics of $\Lambda_b\to\Sigma_c^{(*)}\pi l\nu$. 
Two planes represent the final state hadron and lepton c.m. frame, 
respectively.  The dashed line is the flight line of $s_H$ and $s_L$.
\\
\vskip .3cm

\noindent
Fig.~3
Decay rate distributions for $\Lambda_b\to\Sigma_c\pi l\nu$ over 
(a) $w=v\cdot v^\prime$, (b) pion energy, and (c) lepton energy.
Solid lines are the results of linear $\eta(w)$ (QCD sum rule), 
while the dotted lines are the ones of exponential $\eta(w)$ (large $N_c$).
All the energy scales are in GeV, and the distributions are plotted in
units of $10^{-14}~{\rm or}~10^{-15}~{\rm GeV}$.  
\\
\vskip .3cm

\noindent
Fig.~4
Decay rate distributions for $\Lambda_b\to\Sigma_c^*\pi l\nu$ over 
(a) $w=v\cdot v^\prime$, (b) pion energy, and (c) lepton energy.
Solid lines are the results of linear $\eta(w)$ (QCD sum rule),
while the dotted lines are the ones of exponential $\eta(w)$ (large $N_c$).
All the energy scales are in GeV, and the distributions are plotted in
units of $10^{-14}~{\rm or}~10^{-15}~{\rm GeV}$.

\newpage


\begin{figure}
\vskip 2cm
\begin{center}
\epsfig{file=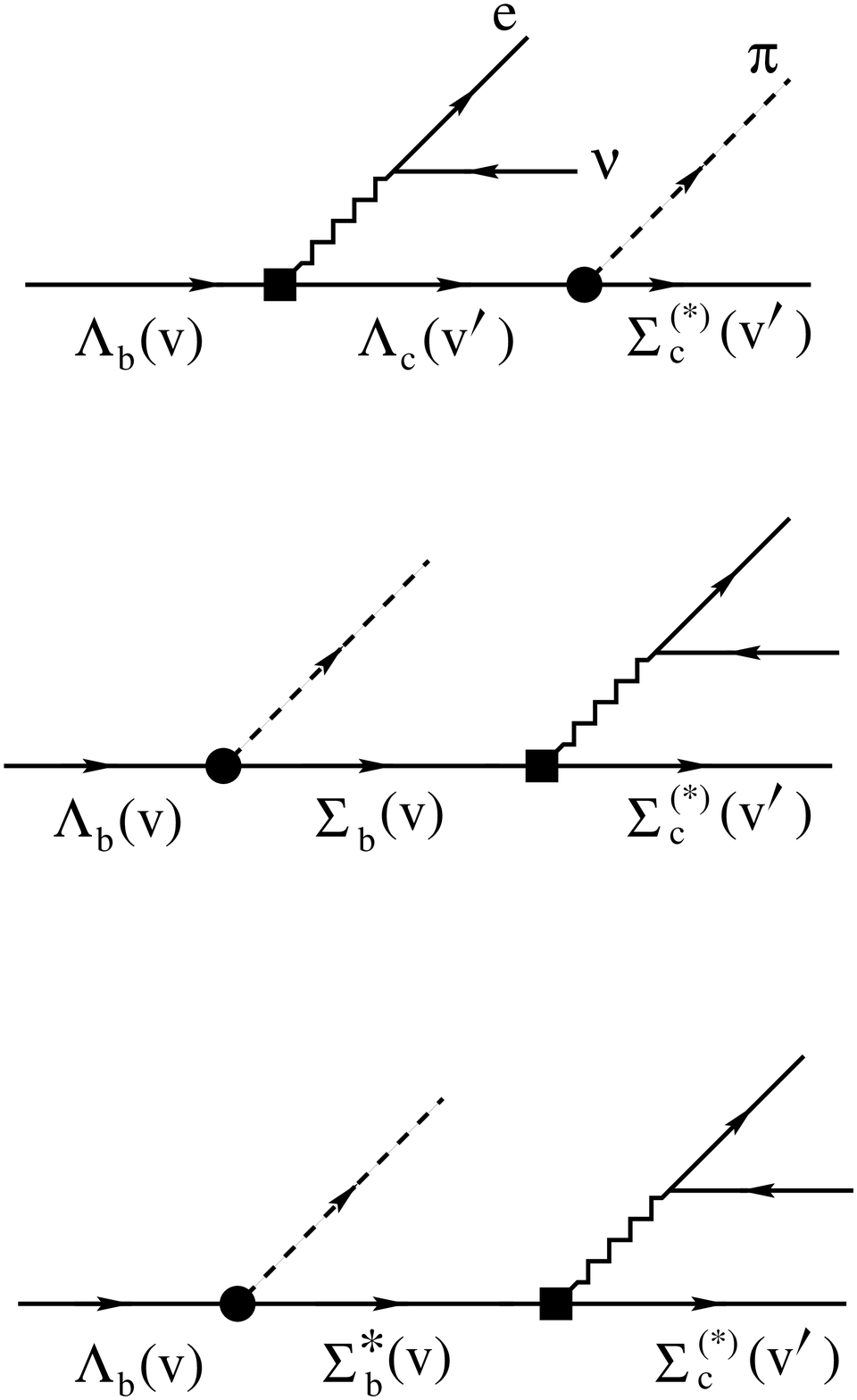, height=10cm}
\end{center}
\caption{}
\end{figure}



\begin{figure}
\vskip 2cm
\begin{center}
\epsfig{file=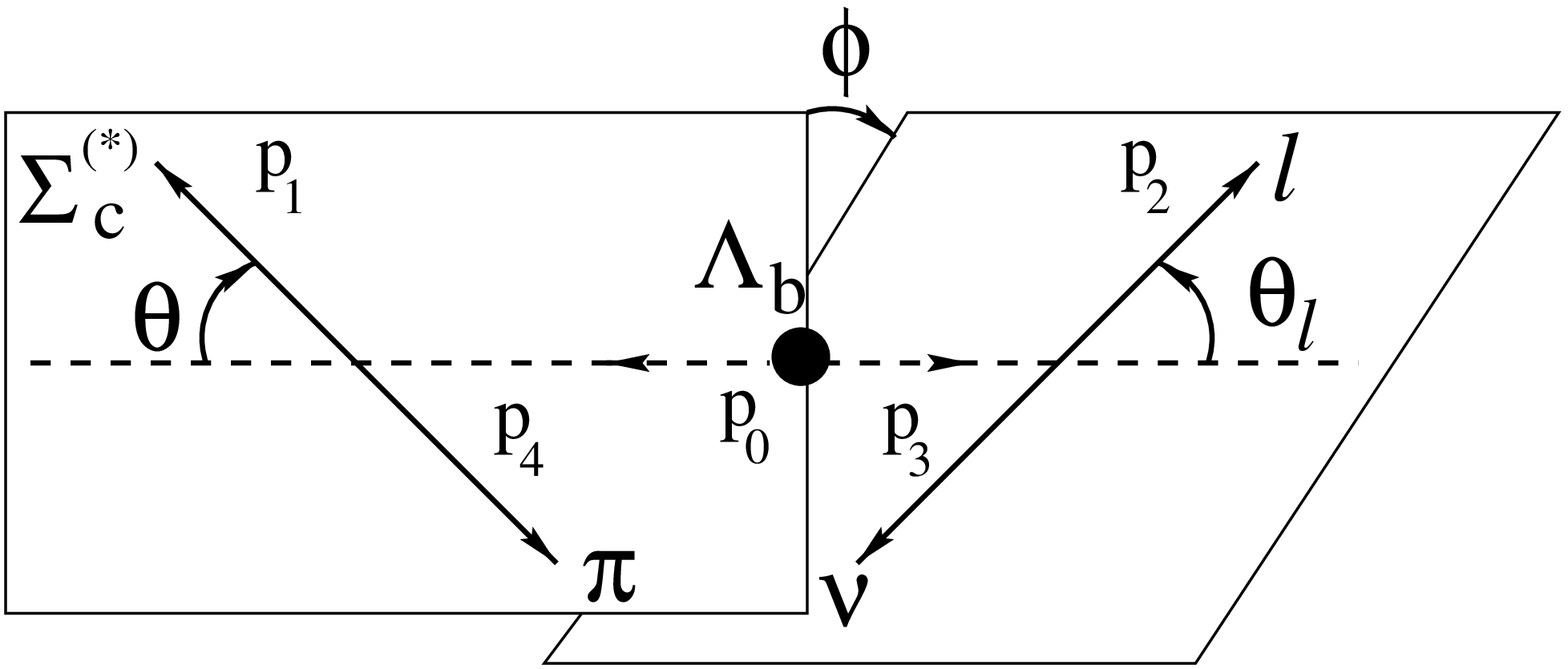, height=5cm}
\end{center}
\caption{}
\end{figure}



\begin{figure}
\begin{tabular}{cc}
\epsfig{file=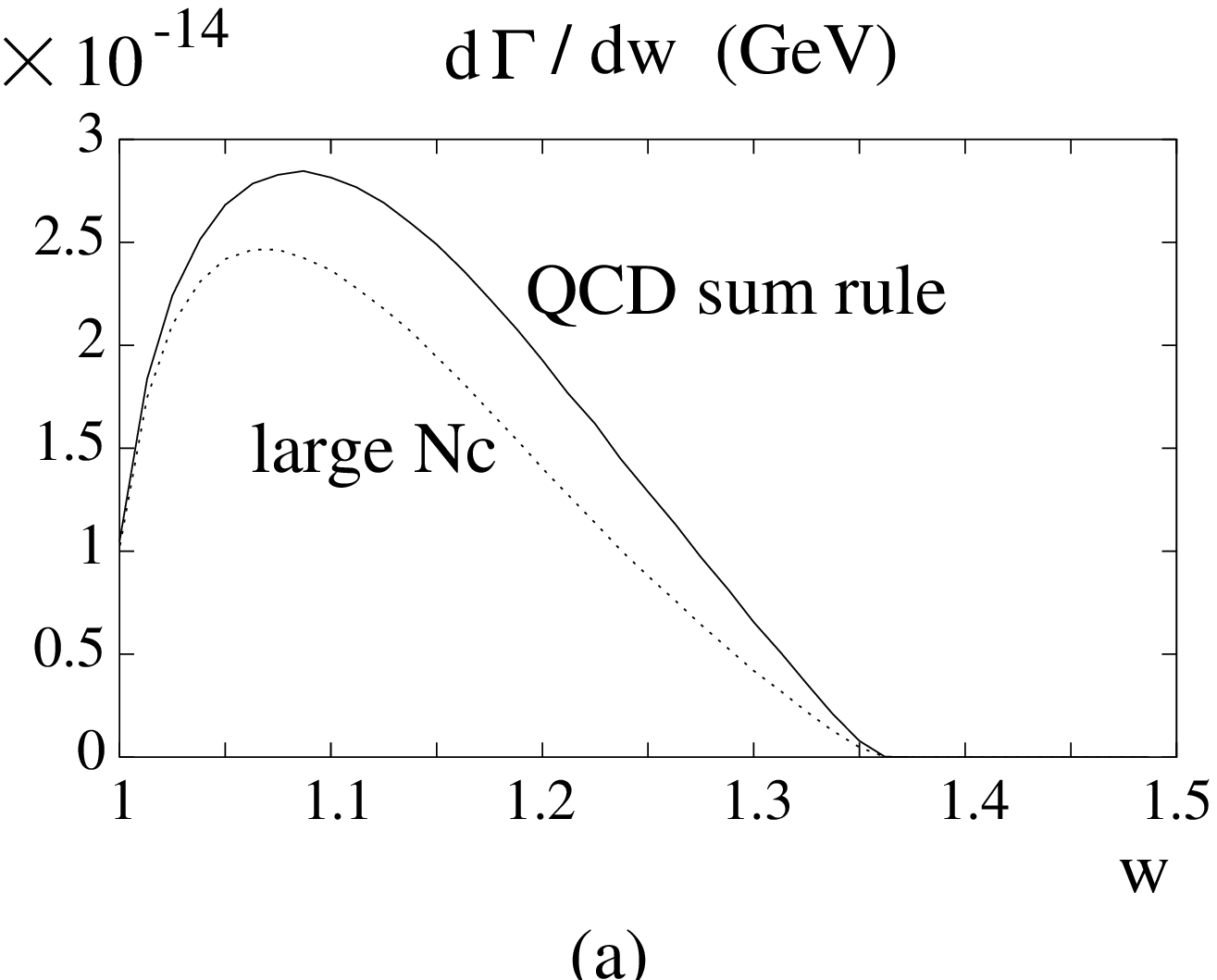, height=5cm} & 
~~~\epsfig{file=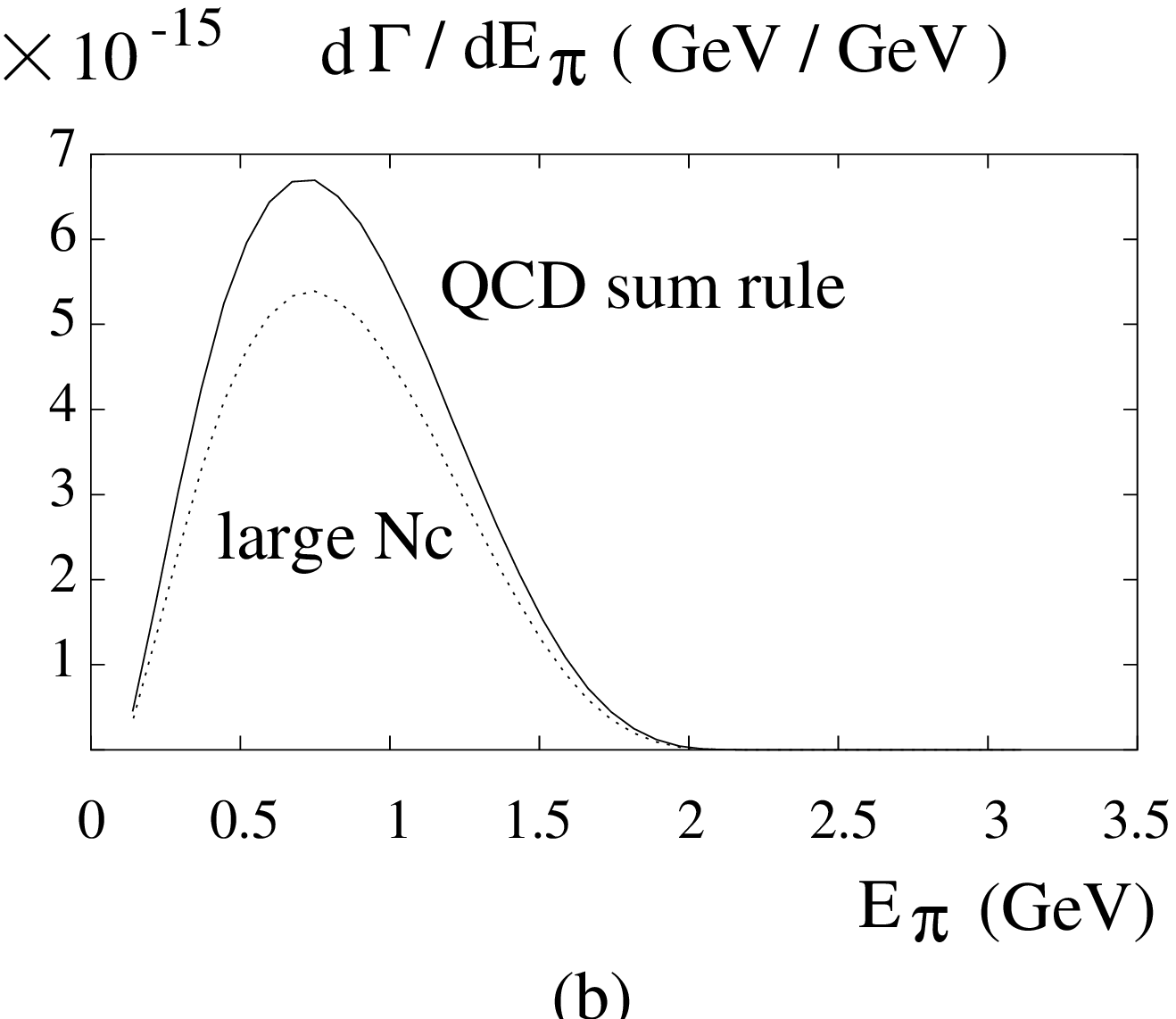, height=5cm}\\[5mm]
\epsfig{file=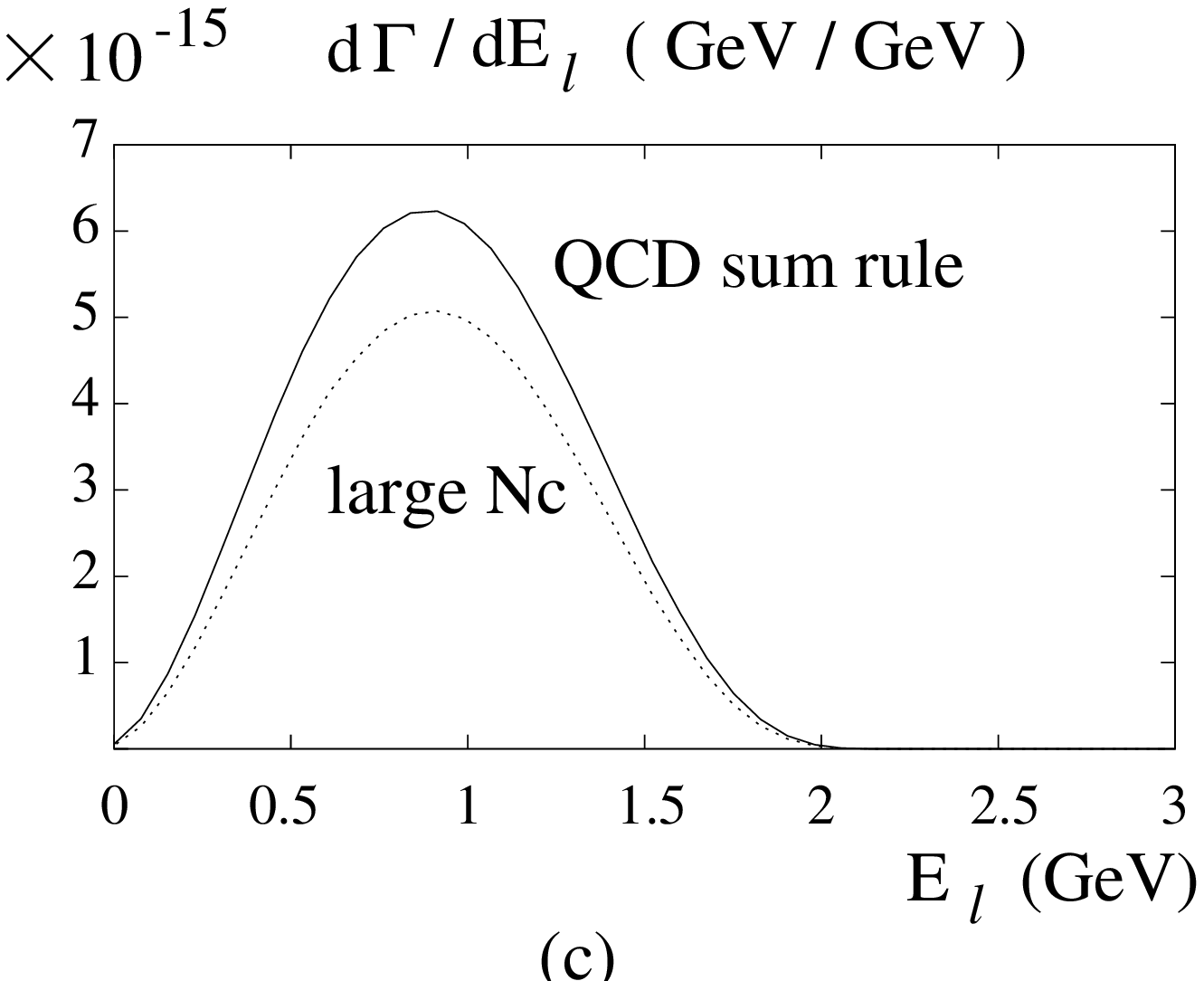, height=5cm}
\end{tabular}
\caption{}
\end{figure} 

\newpage


\begin{figure}
\begin{tabular}{cc}
\epsfig{file=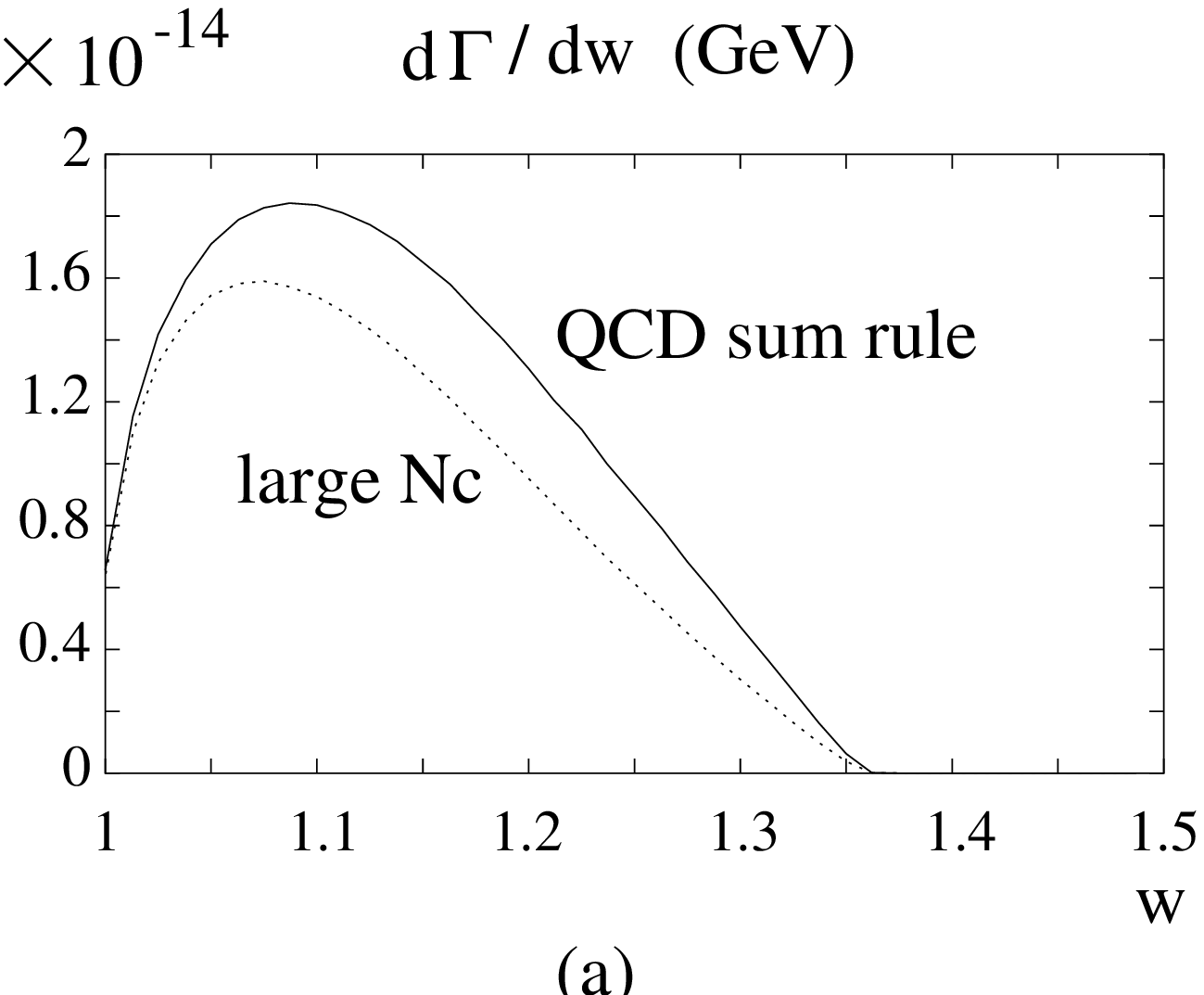, height=5cm} &
~~~\epsfig{file=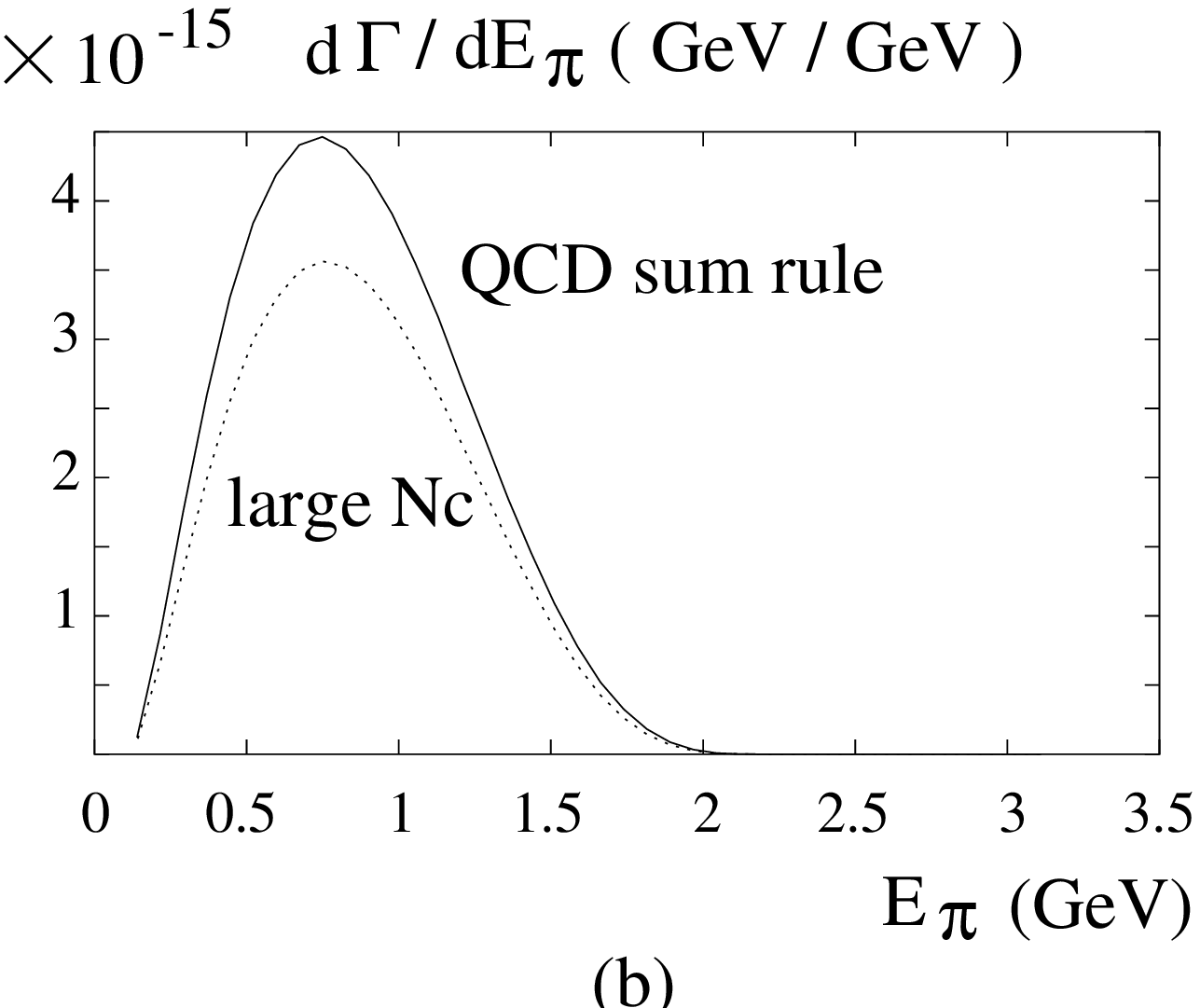, height=5cm}\\[5mm]
\epsfig{file=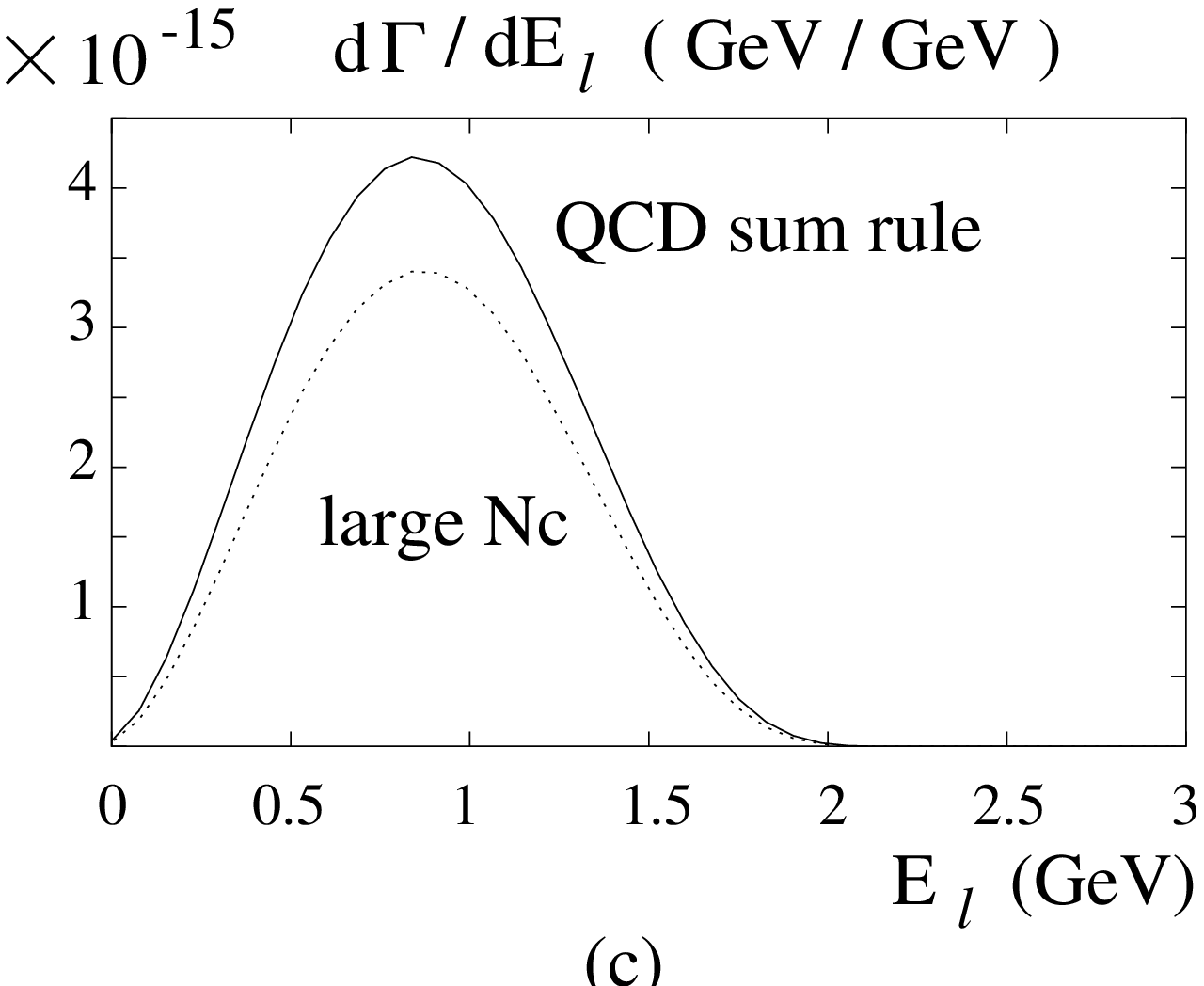, height=5cm}
\end{tabular}
\caption{}
\end{figure}


\end{document}